\documentclass[11pt]{article}

\usepackage{amssymb,amsfonts,amsmath}
\usepackage{afterpage}
\usepackage[final]{graphicx}
\usepackage[letterpaper, total={7in, 9in}]{geometry}
\usepackage{color}
\renewcommand{\baselinestretch}{1.5}

\title{White Matter Network Architecture Guides Direct Electrical Stimulation Through Optimal State Transitions}

\author{Jennifer Stiso$^{1,2}$, Ankit N. Khambhati$^2$, Tommaso Menara$^3$, Ari E. Kahn$^{1,2}$,\\ 
Joel M. Stein$^4$, Sandihitsu R. Das$^5$, Richard Gorniak$^6$,\\ Joseph Tracy$^7$, Brian Litt$^{5,8}$, Kathryn A. Davis$^{5,8}$,\\ Fabio Pasqualetti$^3$, Timothy Lucas$^{8,9}$, \& Danielle S. Bassett$^{2,8,10,11,*}$}

\date{}

\begin{document}

\maketitle

\noindent $^{1}$ Department of Neuroscience, University of Pennsylvania, Philadelphia, PA 19104
 
\noindent $^{2}$ Department of Bioengineering, University of Pennsylvania, Philadelphia, PA 19104
 
\noindent $^{3}$ Department of Mechanical Engineering, University of California, Riverside, CA 92521
 
\noindent $^{4}$ Department of Radiology, Hospital of the University of Pennsylvania, Philadelphia, PA, 19104

\noindent $^{5}$ Department of Neurology, Hospital of the University of Pennsylvania, Philadelphia, PA, 19104
  
\noindent $^{6}$ Department of Radiology, Thomas Jefferson University Hospital, Philadelphia, PA, 19107
 
\noindent $^{7}$ Department of Neurology, Thomas Jefferson University Hospital, Philadelphia, PA, 19107
 
\noindent $^{8}$ Penn Center for Neuroengineering and Therapeutics, University of Pennsylvania, Philadelphia, PA 19104
  
\noindent $^{9}$ Department of Neurosurgery Hospital of the University of Pennsylvania, Philadelphia, PA 19104 

\noindent $^{10}$ Department of Electrical \& Systems Engineering, University of Pennsylvania, Philadelphia, PA 19104

\noindent $^{11}$ Department of Physics \& Astronomy, University of Pennsylvania, Philadelphia, PA 19104

\noindent $^{*}$ To whom correspondence should be addressed: dsb@seas.upenn.edu
 
\clearpage
\newpage

\begin{abstract}
	Electrical brain stimulation is currently being investigated as a potential therapy for neurological disease. However, opportunities to optimize and personalize such therapies are challenged by the fact that the beneficial impact (and potential side effects) of focal stimulation on both neighboring and distant regions is not well understood. Here, we use network control theory to build a formal model of brain network function that makes explicit predictions about how stimulation spreads through the brain's white matter network and influences large-scale dynamics. We test these predictions using combined electrocorticography (ECoG) and diffusion weighted imaging (DWI) data from patients with medically refractory epilepsy undergoing evaluation for resective surgery, and who volunteered to participate in an extensive stimulation regimen. We posit a specific model-based manner in which white matter tracts constrain stimulation, defining its capacity to drive the brain to new states, including states associated with successful memory encoding. In a first validation of our model, we find that the true pattern of white matter tracts can be used to more accurately predict the state transitions induced by direct electrical stimulation than the artificial patterns of a topological or spatial network null model. We then use a targeted optimal control framework to solve for the optimal energy required to drive the brain to a given state. We show that, intuitively, our model predicts larger energy requirements when starting from states that are farther away from a target memory state. We then suggest testable hypotheses about which structural properties will lead to efficient stimulation for improving memory based on energy requirements. We show that the strength and homogeneity of edges between controlled and uncontrolled nodes, as well as the persistent modal controllability of the stimulated region, predict energy requirements. Our work demonstrates that individual white matter architecture plays a vital role in guiding the dynamics of direct electrical stimulation, more generally offering empirical support for the utility of network control theoretic models of brain response to stimulation. 
\end{abstract}

\clearpage
\newpage
\section*{Introduction}

Direct electrical stimulation has demonstrated clinical utility in detecting brain abnormalities during surgery \cite{li2011direct} as well as in mitigating symptoms of epilepsy, essential tremor, and dystonia \cite{Sironi2011,Perlmutter2006,Lozano2013}. Apart from clinical diagnosis and treatment, direct electrical stimulation has also been used to isolate areas responsible for complex higher-order cognitive functions including language \cite{jones2011practical,mani2008evidence}, semantic memory \cite{shimotake2015direct}, and face perception \cite{parvizi2012electrical}. An open and important question is whether such stimulation can be used to enhance cognitive function, and if so, whether stimulation parameters (e.g., intensity and location) can be optimized and personalized based on individual brain anatomy and physiology. While some studies demonstrate enhancements in spatial learning \cite{lee2017stimulation} and memory \cite{Ezzyat2018, Laxton2010,Ezzyat2016,kucewicz2018evidence,Suthana2012} following direct electrical stimulation, others show decrements \cite{Jacobs2016}. Such conflicting evidence is also present in the literature surrounding other types of stimulation, including transcranial magnetic stimulation. Proposed explanations range from variations in stimulation intensity \cite{reichenback2011effects} to individual differences in brain connectivity \cite{downar2014anhedonia}.

\renewcommand{\baselinestretch}{1}
\begin{figure*}
	\centering
	\includegraphics[width=0.95\textwidth]{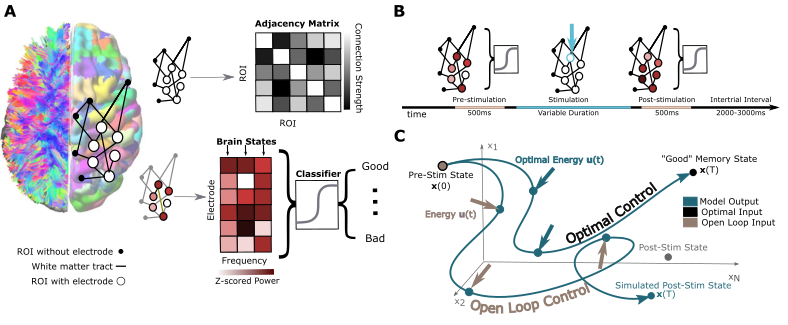}
	\caption{\textbf{Schematic of Methods.} \emph{(A)} Depiction of network construction and definition of brain state. \emph{(Left)} We segment subjects' diffusion weighted imaging data into $N=234$ regions of interest using a Lausanne atlas \cite{Cammoun2012}. We treat each region as a node in a whole-brain network, irrespective of whether or not the region contains an electrode. Edges between nodes represent mean quantitative anisotropy \cite{yeh2013deterministic} along the streamlines connecting them. \emph{(Right, Top)} Practically, we summarize the network in an $N \times N$ adjacency matrix. \emph{(Right, Bottom)} A brain state is defined as the $N \times 1$ vector comprising activity across the $N$ regions. Any element of the vector corresponding to a region with an electrode is defined as the band-limited power of ECoG activity measured by that electrode. Each brain state is also associated with an estimated probability of being in a good memory state, using a previously validated machine learning classifier approach \cite{Ezzyat2016}. \emph{(B)} A schematic of a single stimulation trial. First, ECoG data is collected for 500 ms. Then, stimulation is applied to a given electrode for a variable duration. Finally, ECoG data is again collected after the stimulation. \emph{(C)} A schematic of the open loop and optimal control paradigms. In the open loop design, energy $u(t)$ is applied \emph{in silico} at the stimulation site to the initial, pre-stimulation brain state $x(0)$. The system will travel to some other state $x(T)$ as stipulated by our model of neural dynamics, and we will measure the similarity between that predicted state and the empirically observed post-stimulation state. In the optimal control design, the initial brain state $x(0)$ has some position in space that evolves over time towards a predefined target state $x(T)$. At every time point, we calculate the optimal energy ($u(t)$) required at the stimulating electrode to propel the system to the target state. }
	\label{fig:fig1}
\end{figure*}
\renewcommand{\baselinestretch}{1.5}

A key challenge in circumscribing the utility of stimulation for cognitive enhancement or clinical intervention is the fact that we do not have a fundamental understanding of how an arbitrary stimulation paradigm applied to one brain area alters distributed neural activity in neighboring and distant brain areas \cite{johnson2013neuromodulation,Laxton2010,Lozano2013}. Models of stimulation propagation through brain tissue range in complexity and biophysical realism \cite{McIntyre2004}, from those that only model the region being targeted to those that use finite element models to expand predictions throughout different tissue types \cite{yousif2009investigating}, including both gray matter and white matter \cite{kim2011computational}. Even in the simpler simulations of the effects of stimulation on a local cell population, there are challenges in accounting for the orientation of cells, and the distance from the axon hillock, which can lead to strikingly different circuit behaviors \cite{McIntyre2004}. In the more expansive studies of the effects of stimulation across the brain, it has been noted empirically that minute differences in electrode location can generate substantial differences in which white matter pathways are directly activated \cite{lujan1013brain,riva2014defining}, and that an individual's white matter connectivity can predict successful outcomes of stimulation \cite{Horn2017}. These differences are particularly important in predicting response to therapy, given recent observations that stimulation to white matter may be particularly efficacious in treating depression \cite{riva2013practical} and epilepsy \cite{toprani2013fiber}. Despite these critical observations, a first-principles intuition regarding how the effects of stimulation might depend on the pattern of white matter connectivity present in a single human brain has remained elusive.

Network control theory provides a potentially powerful approach for modeling direct electrical stimulation in humans \cite{tang2017control}. Building on recent advances in physics and engineering, network control theory characterizes a complex system as composed of nodes interconnected by edges \cite{newman2010networks}, and then specifies a model of network dynamics to determine how external input affects the nodes' time-varying activity \cite{Liu2011}. Drawing on canonical results from linear systems and structural controllability \cite{kailath1980linear}, this approach was originally developed in the context of technological, mechanical, and other man-made systems \cite{pasqualetti2014controllability}, but has notable relevance for the study of natural processes from cell signaling \cite{cornelius2013realistic} to gene regulation \cite{zanudo2017structure}. In applying such a theory to the human brain, one first represents the brain as a network of nodes (brain regions) interconnected by structural edges (white matter tracts) \cite{Bassett2017}, and then one posits a model of system dynamics that specifies how control input affects neural dynamics via propagation along the tracts \cite{gu2015controllability}. Formal approaches built on this model address questions of where control points are positioned in the system \cite{gu2015controllability,tang2017developmental,Muldoon2016,wuyan2018benchmarking}, as well as how to define spatiotemporal patterns of control input to move the system along a trajectory from an initial state to a desired final state \cite{gu2017optimal,betzel2016optimally}. Intuitively, these approaches may be particularly useful in probing the effects of stimulation \cite{Muldoon2016} and pharmacogenetic activation or inactivation \cite{Grayson2016} for the purposes of guiding transitions between cognitive states or treating abnormalities of brain network dynamics such as epilepsy \cite{ching2012distributed,ehrens2015closed,Taylor2015}, psychosis \cite{braun2018maps}, or bipolar disorder \cite{jeganathan2018fronto}. However, this intuition has not yet been validated with direct electrical stimulation data.

Here, we posit a simple theory of brain network control, and we test its biological validity and utility in combined electrocorticography (ECoG) and diffusion weighted imaging (DWI) data from patients with medically refractory epilepsy undergoing evaluation for resective surgery. For each subject, we constructed a structural brain network where nodes represented regions of the Lausanne atlas \cite{Cammoun2012} and where edges represented quantitative anisotropy between these regions estimated from diffusion tractography \cite{yeh2013deterministic} (\textbf{Fig.~\ref{fig:fig1}A}). Upon this network, we stipulated a noise-free, linear, continuous-time, and time-invariant model of network dynamics \cite{gu2015controllability,betzel2016optimally,tang2017developmental, gu2017optimal,kim2018role}, from which we built predictions about how regional activity would deviate from its initial state in the presence of exogenous control input to any given node. Using ECoG data acquired from the same individuals during an extensive direct electrical stimulation regimen (\textbf{Fig.~\ref{fig:fig1}B}), we test these theoretical predictions by representing (i) regional activity as an electrode's power in a given frequency band, (ii) the pre-stimulation state as the power prior to stimulation, and (iii) the post-stimulation state as the power after stimulation (\textbf{Fig.~\ref{fig:fig1}C}). After quantifying the relative accuracy of our theoretical predictions, we next use the model to make more specific predictions about the control energy required to optimally guide the brain from a pre-stimulation state to a state associated with good memory encoding. We then test these predictions using subject-level power-based biomarkers of good memory encoding extracted with a multivariate classifier from ECoG data collected during a verbal memory task \cite{Ezzyat2016}. Finally, we investigate how certain topological \cite{kim2018role} and spatial \cite{Roberts2016} properties of a subject's network alter its response to direct electrical stimulation, and we ask whether that response is also modulated by control properties of the area being stimulated \cite{gu2015controllability,Muldoon2016}. Essentially, our study posits and empirically tests a simple theory of brain network control, demonstrating its utility in predicting response to direct electrical stimulation.

\section*{Results}

Our model assumes the time-invariant network dynamics
\begin{equation}
\dot{x}(t) = \mathbf{A} x(t) + \mathbf{B}u(t), \label{eq:1}
\end{equation} where the time-dependent state $x$ is an $N \times 1$ vector ($N=234$) whose $i^{th}$ element gives the band-specific ECoG power in sensor $i$, $\mathbf{A}$ is the $N\times N$ adjacency matrix estimated from DWI data, $\mathbf{B}$ is an $N\times N$ matrix that selects the control set $\mathcal{K} = {u_1, \dots, u_p}$ where $p$ is the number of regions that receive exogenous control input. The input is constant in time and given by $u(t) = \beta \times I\times log(\omega) \times (\Delta t)$, where $I$ is the stimulation amplitude in amperes, $\omega$ is the stimulation frequency in hertz, and $\Delta t$ is the number of simulated samples (here, $950$) divided by the stimulation duration in seconds. The free parameter $\beta$ scales the magnitude of the input (see Materials and Methods). Intuitively, this model formalizes the hypothesis that white matter tracts constrain how stimulation affects brain state.

\subsection*{Predicting Post-Stimulation States by Open Loop Control} 

\begin{figure}[h]
	\centering
	\includegraphics[width=0.4\textwidth]{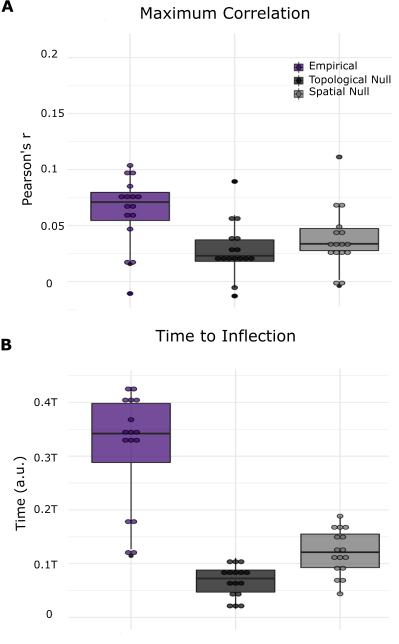}
	\caption{\textbf{Post-Stimulation Brain State Depends on White Matter Network Architecture.} \emph{(A)} Box plots depicting the average signed maximum correlation between the empirically observed post-stimulation state and the predicted post-stimulation state at every time point in the simulated trajectory $x(t)$. \emph{(B)} Box plots depicting the average time to reach the peak magnitude (positive or negative) correlation between the empirically observed post-stimulation state and the theoretically predicted post-stimulation state at every time point in the simulated trajectory $x(t)$. Time is measured in arbitrary units (a.u.). Color indicates theoretical predictions from Eq.1 where $\mathbf{A}$ is (i) the empirical network (purple) estimated from the diffusion imaging data, (ii) the topological null network (dark charcoal), and (iii) the spatial null network (light charcoal).}
	\label{fig:fig2}
\end{figure}

We begin by exercising the model to determine whether our theory accurately predicts changes in brain state induced by direct electrical stimulation. Specifically, we simulate Eq.~\ref{eq:1} to predict how stimulation alone (independent of other ongoing intrinsic dynamics) will alter brain state, given the structural adjacency matrix $\mathbf{A}$ and the initial state $x(0)$ comprised of the ECoG power at every node recorded pre-stimulation ($x_{i} = 1$ if node $i$ is a region without electrodes, and the $z$-scored power otherwise; see Supplement for further details). For each stimulation event, we calculate the Pearson's correlation coefficient between the empirically observed post-stimulation state (an electrode by frequency matrix) and the predicted post-stimulation state at every time point in the simulated trajectory $x(t)$. To measure the capacity of the model simulation to predict the post-stimulation state, we measure the signed maximum correlation achieved across the model simulation time of arbitrary units. Accordingly, we compute a maximum correlation value between the model prediction and the empirically observed post-stimulation state for each stimulation trial ($\mu$ = 0.036 $\pm$ 0.019; \textbf{Fig.~\ref{fig:fig2}A}) and we observe that the mean of the maximum correlation values is significantly greater than zero ($t$-test $N = 16$, $t = 5.83$, $p = 3.31 \times 10^{-5}$). We note that this correlation represents the impact of stimulation alone on linear dynamics, and does not take into account any other incoming stimuli from the surrounding environment, any ongoing cognitive or metabolic processes, nonlinear dependencies, or inter-frequency interactions \cite{Canolty2010,Buzsaki2012,Peterson2017}. Complementing this estimate, we were also interested in the time point (measured in arbitrary units) at which the trial reached its largest magnitude correlation (positive or negative) before decaying towards zero. We observed that the time at which the peak magnitude occurred differed across trials, having a mean of $298 \pm 114$ (\textbf{Fig.~\ref{fig:fig2}B}).

\renewcommand{\baselinestretch}{1}
\renewcommand{\baselinestretch}{1.5}

\renewcommand{\baselinestretch}{1}
\begin{figure}
	\centering
	\includegraphics[width=.95\textwidth]{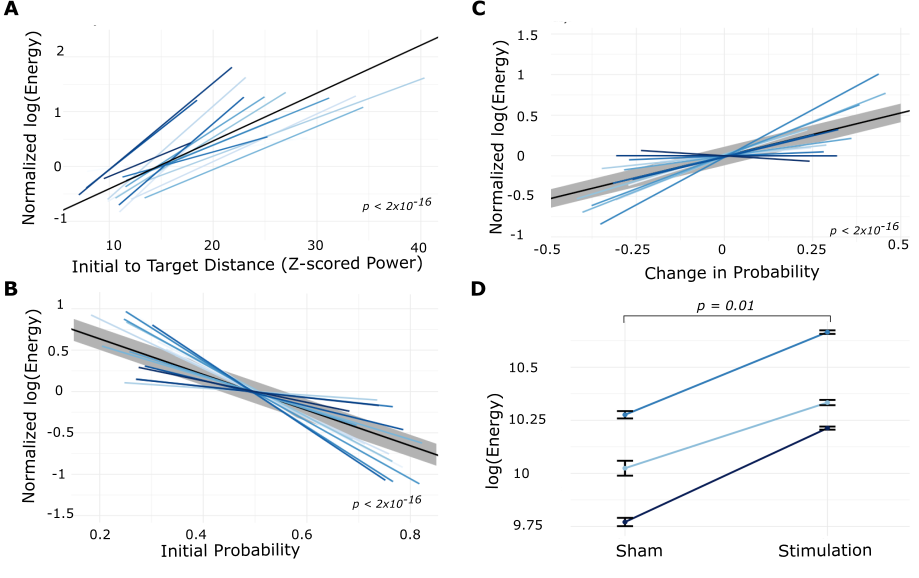}
	\caption{\textbf{Longer-Distance Trajectories Require More Stimulation Energy.} \emph{(A)} The normalized energy required to transition between the initial state and the post-stimulation state, as a function of the Frobenius norm between the initial state and the post-stimulation state. The black solid line represents the best linear fit (with grey representing standard error), and is provided simply as a guide to the eye. Normalization is also done as a visual aid. \emph{(B)} The energy required to transition to a good memory state, as a function of the initial probability of being in a good memory state. \emph{(C)} The energy required to transition to a good memory state as a function of the empirical change in memory state resulting from stimulation. \emph{(D)} In three experimental sessions that included both sham and stimulation trials, we calculated the energy required to reach the post-stimulation state or the post-sham state, rather than a target good memory state. Here we show the difference in energy required for sham state transitions in comparison to stimulation state transitions. Error bars indicate standard error of the mean across trials. Across all four panels, different shades of blue indicate different experimental sessions and subjects.} 
	\label{fig:fig3}
\end{figure}
\renewcommand{\baselinestretch}{1.5}

To determine the influence of network geometry on our model predictions, we compared the empirical observations to those obtained by replacing $\mathbf{A}$ in the simulation with one of two null model networks, each designed to independently remove specific geometric features of the structural network. First for each trial, we constructed a topological null: a randomly rewired network that preserved the edge distribution, number of nodes, and number of edges. Second, we constructed a spatial null: a randomly rewired network that additionally preserved the relationship between edge strength and Euclidean distance. Using a one-way ANOVA, we find a significant difference in maximum correlation values ($F(2,45) = 17.4$, $p = 2.51 \times10^{-6}$), and the time at which the maximum correlation values occur ($F(2,45) = 25.6$, $p = 3.77 \times10^{-8}$). We then performed \emph{post-hoc} analyses and found that the topological null produced significantly weaker maximum correlations between the empirically observed post-stimulation state and the predicted post-stimulation states (paired $t$-test: $N = 16$, $t = 6.58$, uncorrected $p = 8.76\times10^{-6}$), which also peaked significantly earlier in time than the true data ($N = 16$, $t = 7.92$,  uncorrected $p = 7.92\times10^{-7}$). The spatial null model also produced significantly weaker maximum correlations between the empirically observed post-stimulation state and the predicted post-stimulation states (paired $t$-test $N = 16$, $t = 5.83$, uncorrected $p = 3.31\times10^{-5}$), which also occurred significantly earlier in time than that observed in the true data ($N = 16$, $t = 3.84$, uncorrected $p = 1.60\times10^{-3}$). We observed consistent results in individual subjects, across all frequency bands, with different values of $\beta$ and when using a smaller resolution atlas for whole brain parcellation. The only exception was that spatial null models peaked at the same time as empirical graphs in 2 of the 8 frequency bands (see Supplemental Methods). Overall, these observations support the notion that structural connections facilitate a rich repertoire of system dynamics following cortical stimulation, and directly constrain the dynamic propagation of stimulation energy in the human brain in a manner consistent with a simple linear model of network dynamics.

\subsection*{Inducing State Transitions by Optimal Network Control} 

We next sought to use the model to better understand the principles constraining brain state transitions in the service of cognitive function, and their response to exogenous perturbations in the form of direct electrical stimulation. Building on the network dynamics stipulated in \textbf{Eq.~\ref{eq:1}}, we used an optimal control framework to calculate the optimal amount of external input $\mathbf{u}$ to deliver to the control set $K$ containing the stimulating electrode, driving the system from a specific pre-stimulation state towards a target post-stimulation state (\textbf{Fig.~\ref{fig:fig1}C}). This target post-stimulation state was defined as a period with high probability of successfully encoding a memory, and was operationalized using a previously validated classifier constructed from ECoG data from the same subjects during the performance of a verbal memory task \cite{Ezzyat2016} (\textbf{Fig.~\ref{fig:fig1}A}). Specifically, we use a cost function that minimizes both the energy and the difference of the current state from the target state:
\begin{equation}
\underset{u}{min}\int_{0}^{T}(x_{T} - x(t))^{T}\mathbf{S} (x_{T} - x(t)) + \rho ~u(t)^{T} u(t) dt, \label{eq:2}
\end{equation}
where $x_{T}$ is the target state, $\mathbf{S}$ is a diagonal $N\times N$ matrix that selects a subset of states to constrain (here, $\mathbf{S}$ is the identity and all diagonal entries are equal to 1), $\rho$ is the importance of the energy penalty relative to the state penalty, and $T$ is the time allotted for the simulation. Practically, we note that optimizing the cost function in Eq.~\ref{eq:2} necessarily identifies simulated optimal control trajectories from the pre-stimulation state to a good memory state reasonably close to the target (final distance from target $\mu = 0.12 \pm 0.06$) with minimal error (range from $3.65\times10^{-5}$ to $5.19\times10^{-4}$).

We begin by addressing the hypothesis that greater energy should be required to reach the target state when it is farther from the initial state. We operationalize this notion by defining distance in four different ways. First, we define distance as the Frobenius norm of the difference between initial and target states. We fit a linear mixed effects model to the integral of the input squared, or energy (here, $\mathbf{B}u$) in every trial, treating the Frobenius norm distance between initial and final state as a fixed effect, and treating subject as a random effect. We find that the distance between initial and final state is positively related to the energy required for the transition ($\beta = 8.3\times10^{-3}$, $p <  2\times10^{-16}$) (\textbf{Fig.~\ref{fig:fig3}A}). Second, we define distance by the memory capacity in the initial state. We fit a linear mixed effects model to the integral of the input squared in every trial, treating the initial state's probability of successfully encoding a memory as a fixed effect, and treating subject as a random effect. We find that the initial state's probability of successfully encoding a memory is negatively related to the energy required for the transition ($\beta = -0.18$, $p < 2\times10^{-16}$) (\textbf{Fig.~\ref{fig:fig3}B}), suggesting that states that begin closer to the target require less energy to reach the target. Third, we define distance as the observed change in memory state resulting from stimulation. We fit a linear mixed effects model to the input squared in every trial, treating the change in memory state as a fixed effect, and treating subject as a random effect. We find that the change in memory state is positively related to the energy required for the transition ($\beta = 8.3\times10^{-3}$, $p = 2\times10^{-16}$) (\textbf{Fig.~\ref{fig:fig3}C}).

Taken together, this set of results serves as a basic validation that transitions between nearby brain states will generally require less energy than transitions between distant states. This finding holds whether distance is defined in terms of the difference in Frobenius norm between matrices of regional power, or in terms of the estimated probability to support the cognitive process of memory encoding. In specificity analyses, we also determined whether these relationships were expected in appropriate random network null models. We observed that the relationships were significantly attenuated in theoretical predictions from Eq.~1 where $\mathbf{A}$ is either the topological null network ($p = 6.1\times10^{-4}$) or the spatial null network ($p = 0.0017$). Interestingly, we also found that the largest differences between the empirical relationships and those expected in the null networks were observed in the context of biological measures of distance (e.g., initial probability and change in probability), with only modest differences seen in the statistical measure of distance (the Frobenius norm). 

As a fourth and final test of the biological relevance of these findings, we considered sham trials, where no stimulation was delivered, as compared to stimulation trials. Intuitively, we expect that the state that the brain reaches after stimulation is farther away from the initial state than the state that the brain reaches naturally at the conclusion of a sham trial. Consistent with this expectation, we observed that 2 out of the 3 experimental sessions that included sham stimulation displayed significantly larger distances (measured by the Frobenius norm) between pre-and post-stimulation states for stimulation conditions than for sham conditions (permutation test, $p < 6.8\times10^{-3}$). Given this difference, we tested whether more energy would be required to simulate the transition from the initial pre-stimulation state to the post-stimulation state, than from the initial pre-sham state to the post-sham state. We found consistently greater energy for stimulation trials compared to sham trials (paired $t$-test, $p = 0.01$; \textbf{Fig.~\ref{fig:fig3}D}). We further confirmed this finding with a non-parametric permutation test assessing differences in the distribution of energy values across trials for sham conditions and the distribution of energy values across trials for stimulation conditions (permutation test, $p < 2\times10^{-16}$ for all subjects). These observations support the notion that transitions between nearby brain states occur without stimulation (sham) and require little predicted energy, whereas transitions between distant brain states occur with stimulation and require greater predicted energy.

\subsection*{The Role of Network Topology on Stimulation-Based Control}

\begin{figure}
	\centering
	\includegraphics[width=0.5\textwidth]{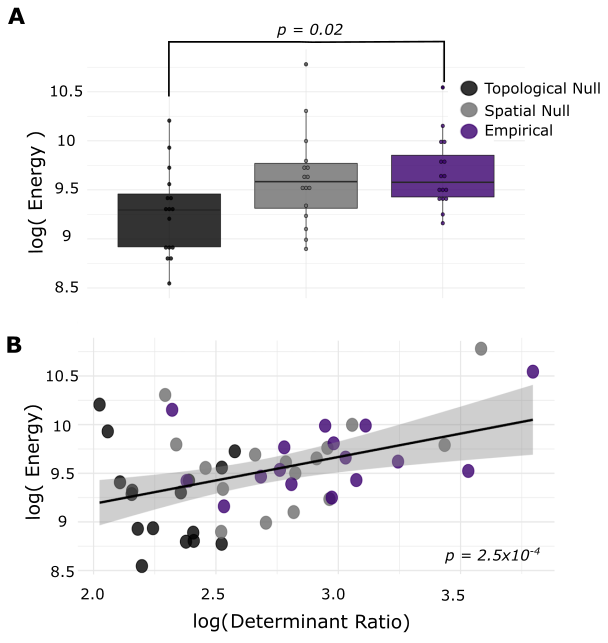}
	\caption{\textbf{Topological and Spatial Constraints on the Energy Required for Stimulation-Based Control.} \emph{(A)} Average input energy required for each transition from the pre-stimulation state to a good memory state, as theoretically predicted from Eq.~1 where $\mathbf{A}$ is (i) the empirical network (purple) estimated from the diffusion imaging data, (ii) the topological null network (dark charcoal), and (iii) the spatial null network (light charcoal). \emph{(B)} The relationship between the determinant ratio and the energy required for the transition from the pre-stimulation state to a good memory state. Note: The color scheme is identical to that used in panel \emph{(A)}.}
	\label{fig:fig4}
\end{figure}

While it is natural to posit that the distance between brain states is an important constraint on the ease of a state transition, there are other important principles that are also likely to play a critical role. Paramount among them is the architecture of the network available for the transmission of control signals. We therefore now turn to the question of which features of the network predict the amount of energy required for each transition from the pre-stimulation state to a good memory state. To address this question, we considered the empirical networks as well as the topological and spatial null model networks discussed earlier. We find that the optimal control input energy required for these state transitions differs across network types (one-way ANOVA $F(2,75) = 4.00$, $p = 0.03$). In \emph{post-hoc} testing, we found that the optimal control energy was significantly different between the empirical network and the topological null network (two-tailed $t$-test: $t = -2.6$, $p = 0.01$) (\textbf{Fig.~\ref{fig:fig4}A}), but not between the empirical network and the spatial null network ($p>0.05$). This observation suggests that the spatial embedding that characterizes both the real network and the spatial null network may increase the difficulty of control. In supplemental analyses, we test two additional spatially embedded null models that further preserve degree distribution and strength sequence, and we find similar average energies to the empirical and spatial null models discussed here (see Supplement). We hypothesized that the difference in optimal control energy could be mechanistically explained by the determinant ratio, a recently proposed metric quantifying the trade-off between connection strength (facilitating control) and connection homogeneity (hampering control) \cite{kim2018role}. Intuitively, a network with a high determinant ratio will have weak, homogenous connections between the control nodes and nodes being controlled. We found that across all networks the determinant ratio explains a significant amount of variance in energy after accounting for network type (linear mixed effects model with network type and determinant ratio as fixed effects: $\chi^{2} = 12.2$, $p = 4.7\times10^{-4}$) (\textbf{Fig.~\ref{fig:fig4}B}). These results support the notion that spatial embedding could impose energy barriers by compromising the trade-off between the strength and homogeneity of connections emanating from the stimulating electrode.

\renewcommand{\baselinestretch}{1}
\renewcommand{\baselinestretch}{1.5}

\subsection*{Characteristics of Efficient Regional Controllers}

\begin{figure}
	\centering
	\includegraphics[width=0.95\textwidth]{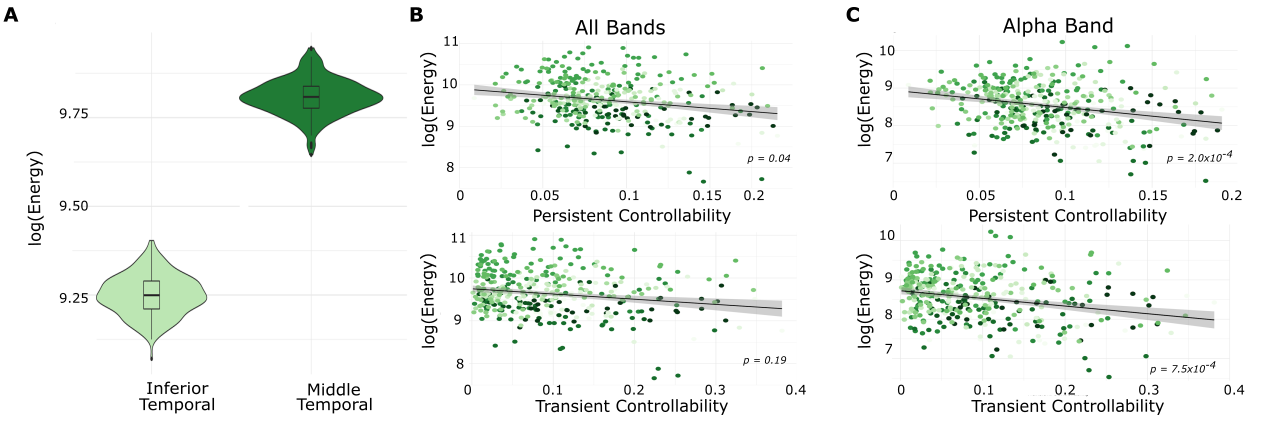}
	\caption{\textbf{Role of Local Topology Around the Region Being Stimulated.} \emph{(A)} Transitions from the observed initial state to a good memory state required significantly greater energy when affected by the middle temporal sensors than when affected by the inferior temporal sensors. \emph{(B)} Relationship between persistent (\emph{top}) or transient (\emph{bottom}) controllability of the stimulated region and the energy predicted from optimal transitions from the initial state to a good memory state.  We only allow energy to be injected into a single electrode-containing region, and we consider a broadband state matrix. \emph{(C)} As in panel \emph{(B)} but when considering the $\alpha$ band state vector only.}
	\label{fig:fig5}
\end{figure}

Thus far, we have seen that the distance of the state transition and the architecture of the network available for the transmission of control signals both impact the energy required. However, neither of these factors addresses the potential importance of anatomical characteristics specific to the region being stimulated. Such regional effects are salient in the one subject in our patient sample who had multiple empirical stimulation sites spanning the same number of ROIs. In this patient, we found that transitions from the observed initial state to a good memory state required significantly greater energy when stimulation was delivered to electrodes in the middle temporal region than when stimulation was delivered to the inferior temporal region (permutation test, $p < 2\times10^{-16}$) (\textbf{Fig.~\ref{fig:fig5}A}). We hypothesized that this sensitivity to anatomical location could be mechanistically explained by regional persistent and transient modal controllability, which quantify the degree to which specific eigenmodes of the network's dynamics can be influenced by input applied to that region. Energetic input to nodes with high persistent controllability will result in large perturbations to slowly decaying modes of the system, while energetic input to nodes with high transient controllability will result in large perturbations to quickly decaying modes of the system. 

\renewcommand{\baselinestretch}{1}
\renewcommand{\baselinestretch}{1.5}

To test our hypothesis, we simulated optimal trajectories from the initial state to a good memory state while only allowing energy to be injected into a single electrode-containing region (irrespective of whether or not empirical stimulation was applied there). We then compared the energy predicted from these simulations to the regional controllability. We found a significant relationship between persistent (but not transient) modal controllability of the region being stimulated and the input energy of the state transition (linear mixed effects model accounting for subject: persistent controllability $\chi^{2} = 3.89$, $p = 0.049$, transient controllability $\chi^{2} = 1.69$, $p = 0.19$) (\textbf{Fig. ~\ref{fig:fig5}B}). We note that the strength of the region being stimulated was not a significant predictor of energy (linear mixed effects model $\chi^{2} = 3.5$, $p = 0.061$). Notably, we found that the broad-band effect was heavily driven by the $\alpha$ band (linear mixed effects model: persistent controllability $\chi^{2} = 13.8$, $p = 2.0\times10^{-4}$, transient controllability $\chi^{2} = 11.4$, $p = 7.5\times10^{-4}$; Bonferroni corrected for multiple comparisons across frequency bands) (\textbf{Fig.~\ref{fig:fig5}C}). Additionally, in the one subject that had two empirical stimulation locations, we observed that the middle temporal stimulation site with larger energy requirements had smaller persistent controllability (0.058) than the inferior temporal site with smaller energy (0.072). These findings suggest that the local white matter architecture of stimulated regions can support the selective control of slowly damping dynamics.

\subsection*{Effective Prediction of Energy Requirements}

\begin{figure}
	\centering
	\includegraphics[width = .55\textwidth]{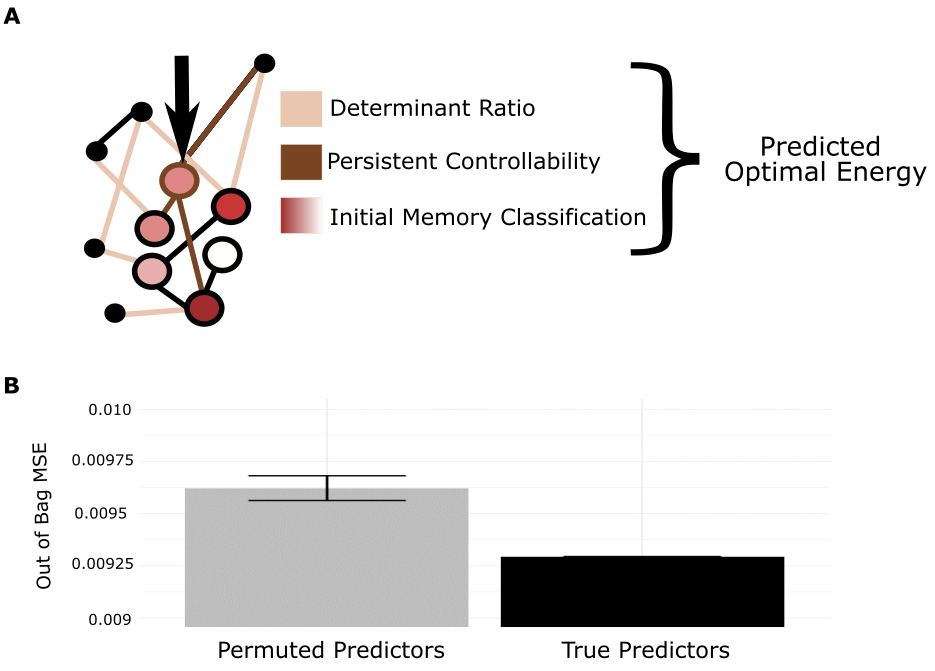}
	\caption{\textbf{Network Topology and Brain State Predict Energy Requirements.} \emph{(A)} Schematic of the three topology and state features included in the random forest model that we built to predict energy requirements. Network level effects (tan) are captured by the determinant ratio, regional effects (brown) are captured by persistent controllability, and state-dependent effects (red) are captured by the initial memory state. \emph{(B)} Comparison of the out-of-bag mean squared error for a model where each subject's determinant ratio, persistent controllability, and initial memory state are used to predict their required energy. We compared the performance of this model to the performance of a distribution of 1000 models in which the association between energy values and predictors was permuted uniformly at random. }
	\label{fig:fig6}
\end{figure}

In the previous section, we presented a series of analyses with the goal of elucidating what aspects of brain state and white matter connectivity affect the energy requirements predicted by our model, in an effort to better understand the network wide effects of direct electrical stimulation. Here, we conclude by synthesizing these results into a single model to predict the energy requirements of a stimulation paradigm, given the persistent controllability of the region to be stimulated, the determinant ratio of the network to be controlled, and the probability of encoding a memory at the time of stimulation (Fig\ref{fig:fig6}A). We fit a random forest model to predict energy given these inputs from our data, and we compared the performance of this model to the performance of a distribution of 1000 models in which the association between energy values and predictors was permuted uniformly at random. We found that our model had an out-of-bag mean squared error of $9.28\times 10^{-3}$, substantially lower than the null distribution ($\mu = 9.62\times10^{-3} \pm 2.97\times10^{-5}$). We also found that our model explained $93.2\%$ of the variance in the predicted energy of the state transition. Random forest models also produce a measure of variable importance, which represents the degree to which including these variables tends to reduce the prediction error. We found that the determinant ratio was the most important (increased node purity = 627), followed by the persistent controllability (320), followed by the initial probability of encoding a memory (23.0). Broadly, these results suggest that the energy requirements for a specific state transition can be accurately predicted given simple features of the connectome and the current brain state.

\renewcommand{\baselinestretch}{1}
\renewcommand{\baselinestretch}{1.5}

\section*{Discussion}

While direct electrical stimulation has great therapeutic potential, its optimization and personalization remains challenging, in part due to a lack of understanding of how focal stimulation impacts both neighboring and distant regions. Here use network control theory to test the hypothesis that the effect of direct electrical stimulation on brain dynamics is constrained by an individual's white matter connectivity. By stipulating a simplified noise-free, linear, continuous-time, and time-invariant model of neural dynamics, we demonstrate that time-varying changes in the pattern of ECoG power across brain regions is better predicted by an individual's true white matter connectivity than either topological or spatial network null models. We build on this observation by positing a model for brain state transitions in which the energy required for the state transition is minimized, as is the length of the trajectory through the available state space. We demonstrate that transitions between more distant states are predicted to require greater energy than transitions between nearby states; these results are particularly salient when distance is defined based on differences in the probability with which a cross-regional pattern of ECoG power supports memory encoding. In addition to the distance between initial and target states, we also find that regional and global characteristics of the network topology predict the energy required for the state transition: networks with smaller determinant ratios (stronger, less homogeneous connections), and stimulation regions with higher persistent controllability, tend to demand less energy. Finally, we demonstrate that these two topological features in combination with the initial brain state explain 93\% of the variance in required energy across subjects. Overall, our study supports the notion that control theoretic models of brain network dynamics provide biologically grounded, individualized hypotheses of response to direct electrical stimulation by accounting for how white matter connections constrain state transitions.

\subsection*{A Role for Control Theory in Modern Neuroscience}

Developing theories, models, and methods for the control of neural systems is not a new goal in neuroscience. Whether in support of basic science (e.g., seminal experiments from Hodgkin and Huxley) or in support of clinical therapies (e.g., technological development in brain-machine interfaces or deep brain stimulation), efforts to control neural activity have produced a plethora of experimental tools with varying levels of complexity \cite{schiff2011neural}. Building on these empirical advances, the development of a theory for control in neuroscience is a logical next step. Network control theory is one particularly promising option. In assimilating brain state and connectivity in a mathematical model \cite{schiff2011neural}, network control theory offers a first-principles approach to modeling neural dynamics, predicting its response to perturbations, and optimizing those perturbations to produce a desired outcome. In cellular neuroscience, network control theory has offered predictions of the functional role of individual neurons in \emph{C. elegans}, and those predictions have been validated by perturbative experiments \cite{Yan2017}.  While the theory has also offered predictions in humans \cite{gu2015controllability,Muldoon2016,ching2012distributed,Taylor2015,jeganathan2018fronto}, these predictions have not been validated in accompanying perturbative experiments. Here we address this gap by examining the utility of network control theory in predicting empirically recorded brain states, and by validating the fundamental assumption that state transitions are constrained by an individual's white matter connectivity. The work provides theoretical support for emerging empirical observations that structural connectivity can predict the behavioral effects of stimulation \cite{Horn2017}, thus constituting an important first step in establishing the promise and utility of control theoretic models of brain stimulation.

\subsection*{The Principle of Optimal Control in Brain State Transitions} 

By positing a model for optimal brain state transitions, we relate expected energy expenditures to a change in the probability with which a pattern of ECoG power is associated with good memory encoding, directly relating the theory to a desired behavioral feature. This portion of the investigation was made possible by an important modeling advance addressing the challenge of simulating a trajectory whose control is dominated by a single node: the stimulating electrode. This type of control is an intuitive way to model stimulation, where you only want to capture changes resulting from a single input source. However, prior work has demonstrated that while the brain is theoretically controllable from a single point, the amount of energy required can be so large as to make the control strategy impractical \cite{gu2015controllability}. Here we extend prior models of optimal control \cite{betzel2016optimally, gu2017optimal} by relaxing the input matrix $\mathbf{B}$ such that it allows large input to stimulated regions, but also allows small, randomly generated amounts of input at other nodes in the network. Practically, this approach greatly lowers the error of the calculation and also produces narrowly distributed trajectories for the same inputs (see Supplemental Methods). 

\subsection*{Topological Influencers of Control} 

Beyond the distance of the state transition, we found that both local and global features of the network topology were important predictors of control energy. In line with previous work investigating controllability radii \cite{Menara2018}, energy requirements were lower for randomly rewired networks. Both empirical and topological graphs share the common feature of modularity \cite{Chen2013}, which is destroyed in random topological null models \cite{Roberts2016}. Prior theoretical work has demonstrated that modularity is one way in which to decrease the energy of control by decreasing the determinant ratio, a quantification of the relationship between the strength and heterogeneity of direct connections from the controlling node to others \cite{kim2018role}. Here we confirmed that the determinant ratio accurately predicted the required energy, while leaving a small amount of variance unexplained. We expected that this unexplained variance could be somewhat accounted for by features of the local network topology surrounding the stimulated node \cite{tang2017developmental}. Consistent with our expectation, we found that persistent controllability was the only significant predictor of energy across all frequency bands, indicating a specific role of slow modes in these state transitions. The effect was particularly salient in the $\alpha$-band, whose role in memory encoding is well-known \cite{Fell2011,Buzsaki2013}. 

\subsection*{Methodological Considerations}

\subsubsection*{Primary Data} As with any model of complex biological systems, our results must be interpreted in the context of the underlying data. First, we note that DWI data provides an incomplete picture of white matter organization, and even state-of-the-art tractography algorithms can identify spurious connections \cite{Thomas}. As higher resolution imaging, reconstruction, and tractography methods emerge, it will be important to replicate the results we report here. Second, while ECoG data provides high temporal resolution, it is collected from patients with epilepsy and results might not generalize to a healthy population \cite{Parvizi2017}. However, it is worth noting that recent work has shown that tissue damage resulting from recurrent seizures can be minimal \cite{Rossini2017}, and most electrodes are not placed in epileptic tissue \cite{Parvizi2017}. Nevertheless, this population can display atypical physiological signatures of memory \cite{Glowinski1973}, as well as atypical white matter connectivity \cite{Gross2006}. It will be important in future to extend this work to non-invasive techniques accessible to healthy individuals. 

\subsubsection*{Modelling Assumptions} Our results must also be interpreted in light of model assumptions. First, we note that our model assumes linear network dynamics. While the brain is not a linear system, such simplified approximations can predict features of fMRI data \cite{Honey}, predict the control response of nonlinear systems of coupled oscillators \cite{Muldoon}, and more generally provide enhanced interpretability over nonlinear models. Nevertheless, considering control in nonlinear models of neural dynamics will constitute an important next step. Second, we consider a relaxed input matrix to ensure that state transitions are primarily influenced by the set of stimulating electrodes and to a lesser extent non-stimulating electrodes. This choice is not a true representation of single point control, but instead reflects the fact that the system is constantly modulated by endogenous sources \cite{gu2017optimal,betzel2016optimally}. Lastly, our model uses a time-invariant connectivity matrix. While DWI data is relatively stable over short time-scales, repeated stimulation can result in dynamic changes in plasticity \cite{Malenka2004} that are not captured here.

\subsubsection*{Defining Brain States} In our model, a brain state represents the $z$-scored power across electrodes in eight logarithmically spaced frequency bands from 1 to 200 Hz. This choice was guided by (i) the goal of maintaining consistency with the brain states on which the memory classifier was trained, and (ii) the fact that power spectra are well-documented behavioral analogs for memory \cite{Ezzyat2016, Fell2011, Buzsaki2013}. Yet, since many power calculations require convolution with a sine wave, power is insensitive to non-sinusoidal and phase-dependent features of the signal \cite{Schalk2017,Cole2017,Vinck2011}. It would be interesting in future to explore transitions in other state spaces, such as instantaneous voltage \cite{Schalk2017}. Lastly, it is important to note that our algorithm controls each frequency band independently, although incorporating inter-frequency coupling \cite{Peterson2017,Bonnefond,Canolty2010} could be an interesting direction for future work.

\subsection*{Conclusions and Future Directions} Our study begins to explore the role of white matter connectivity in guiding direct electrical stimulation, with the goal of driving brain dynamics towards states with a high probability of memory encoding. We demonstrate that our model of targeted direct electrical stimulation tracks well with biological intuitions, and is influenced by both regional and global topological properties of underlying white matter connectivity. Overall, we have shown that our control theoretic model is a promising method that has potential to inform hypotheses about the outcome of direct electrical stimulation.

\section*{Materials and Methods}

\subsection*{Data Acquisition and Preprocessing}

Electrocorticography data were collected from eleven subjects (age 32 +/- 100 years, 63.6\% male and 36.4\% female) at Thomas Jefferson University Hospital and the Hospital of the University of Pennsylvania as part of a multi-center project designed to assess the effects of electrical stimulation on memory-related brain function. The research protocol was approved by the institutional review board (IRB) at each hospital and informed consent in writing was obtained from each participant. Electrophysiological data were collected from electrodes implanted subdurally on the cortical surface as well as deep within the brain parenchyma. In each case, the clinical team determined the placement of the electrodes to best localize epileptogenic regions. Subdural contacts were arranged in both strip and grid configurations with an inter-contact spacing of 10 mm. Depth electrodes had 8-12 contacts per electrode, with 3.5 mm spacing. In our model, a brain state represented the $z$-scored power across electrodes in eight logarithmically spaced frequency bands from 1 to 200 Hz.

Diffusion volumes were skull-stripped using FSL's BET, v5.0.10. Volumes were subsequently corrected for eddy currents and motion using FSL's EDDY tool, v.5.0.10 \cite{Andersson2016}). Anatomical scans were processed with FreeSurfer (http://surfer.nmr.mgh.harvard.edu/) v6.0.0. Surface reconstructions were used to generate subject-specific parcellations based on the Lausanne atlas from the Connectome Mapper Toolbox \cite{Daducci2012}. Each parcel was then individually warped into the subject's diffusion space. Using DSI-Studio (http://dsi-studio.labsolver.org), orientation density functions (ODFs) within each voxel were reconstructed from the corrected scans using GQI \cite{yeh2013deterministic}. We then used the reconstructed ODFs to perform a whole-brain deterministic tractography using the derived QA values in DSI-Studio \cite{yeh2013deterministic}. We generated 1,000,000 streamlines per subject, with a maximum turning angle of 35 degrees and a maximum length of 500mm \cite{Cieslak2014}. We held the number of streamlines between participants constant \cite{Griffa2013}. 

\subsection*{Memory State Classification}
Prior to collecting the stimulation data used in this study, non-stimulation ECoG data was collected from each subject as they were performing a verbal memory task. This data was used to train a memory classifier, built on the spectral power averaged across the time dimension for each word-encoding epoch. Each subject's personalized classifier was then used to return a probability of being in a good memory state for each pre- and post-stimulation recording (for further details, see \cite{Ezzyat2016}). We used this information to define the target state for our simulations as the average of the top 5\% of states with the largest probabilities associated with them. The threshold of 5\% was chosen as the smallest threshold that reliably included a sufficient average number of trials, with the goal of only selecting for memory-relevant features and not noise. The probabilities associated with these final target states ranged from 0.61 to 0.74.

\subsection*{Post-Stimulation State Correlations}
We simulated stimulation to a given region in the Lausanne atlas from the observed pre-stimulation state ($x(i)$ is the $z$-scored power if $i$ is a region with an electrode, $x(i) = 1$ otherwise). We then calculated the two-dimensional Pearson's correlation coefficient between the empirically observed post-stimulation state and the predicted post-stimulation state at time points $t = 5$ to $t = T$ in the simulated trajectory $x(t)$. The time points $t < 5$ were excluded to prevent the initial state from being considered as the peak. We calculated two statistics of interest: the maximum correlation reached and the time at which the largest magnitude (positive or negative) correlation occurred. 

\subsection*{Null Models}
We compared the empirically observed values -- of the maximum correlation reached and the time at which the largest magnitude correlation occurred -- to those expected under two null models: (i) a topological null model that preserved only the number of edges and their total strength, and (ii) a spatially embedded null model that also preserved the relationship between edge strength and edge distance. Instantiations of the topological null model were generated using the Brain Connectivity Toolbox \cite{Rubinov2010}. The rewiring algorithm begins by randomly choosing two pairs of edges ($i \rightarrow j$ and $k \rightarrow l$) and continues by swapping their origin and termination points ($i \rightarrow k$ and $j \rightarrow l$). Here, we performed $2 \times 10^{4}$ bidirectional edge swaps per network. Instantiations of the spatially embedded model were generated using code from \cite{Roberts2016}. The rewiring algorithm begins by calculating the Euclidean distance between the average coordinates of all regions in the Lausanne atlas, and continues by removing the effect of distance on the mean and variance of the edge weights, randomly rewiring, and then adding the effect of distance back to the newly rewired graph. For both topological and spatial null model analyses, a new random graph was generated for every trial (minimum number of trials was 192). Null models were created from the stabilized rather than raw versions of the structural matrices, and -- in the optimal control analyses -- were also scaled by a parameter $\gamma$ to reduce the error of the calculation (see Supplemental Materials).

\subsection*{Optimal Network Control} 

To quantify the ease of controlling the system, we calculated a single measure of energy for every trajectory. We used a measure of total input energy that incorporated the weights of $\mathbf{B}$ in addition to the input $\mathbf{u}$ because the entries of $\mathbf{B}$ were graded:
\begin{equation}
E_{\kappa,\mathbf{x}_0\mathbf{x}_T} = \int_{0}^{T}||\mathbf{B_{\kappa}u}_{\mathbf{x}_0\mathbf{x}_T}||_{2}^{2} \mathrm{d}t .
\end{equation}
More specifically, rather than being characterized by binary state values, regions without electrodes were given a value of approximately $5\times10^{-5}$ at their corresponding diagonal entry in $\mathbf{B}$. This additional input ensured that the calculation of optimal energy was computationally tractable (which is not the case for input applied to a very small control set), but also necessitated the incorporation of $\mathbf{B}$ into the calculation of energy for a more representative estimate. 

Trajectories were simulated for each frequency band, and these trajectories were combined into a single state matrix for all analyses, unless otherwise specified (e.g., as in Fig.~\ref{fig:fig5}C and in some figures in the Supplementary Materials). More specifically, comparisons of brain state were calculated as the two-dimensional Pearson's correlation coefficient between simulated region-by-frequency matrices and empirical region-by-frequency matrices (Fig.~\ref{fig:fig2}). Only regions with electrodes were included in correlations, as they were the only regions with initial state measurements. Energy in all optimal control analyses was calculated in each band independently, and then summarized in a region-by-frequency matrix at each time point (Fig.~\ref{fig:fig3}--\ref{fig:fig6}). A single measure of energy for a trial was calculated by integrating the Frobenius norm of the energy matrix over time.

\subsection*{Network Statistics} To probe the role of graph architecture in the energy required for optimal control trajectories, we calculated the determinant ratio, which is defined as the ratio of the strength to the homogeneity of the connections between the first degree driver (anything with a non-zero entry in $\mathbf{B}$) and the non-driver (anything with a zero entry in $\mathbf{B}$) \cite{kim2018role}. This metric was derived assuming that a system has a greater number of driver nodes than non-driver nodes, and that the initial and final states are distributed around zero; importantly, both assumptions are accurate for our simulations. Quantitatively, the trade-off between strength and homogeneity is embodied in the ratio between the determinant of the Gram matrix of all driver to non-driver connections, and the determinant of that same matrix with each non-driver node removed iteratively. The gram matrix here is the inner product of the vectors giving connections from driver nodes to and non-driver nodes. More specifically, if $C$ is the Gram matrix of all driver to non-driver connections, and $C_{k}$ is the matrix of all connections from driver nodes to all but the $k^{\mathrm{th}}$ non-driver node, the determinant ratio is defined by $N^{-1}\sum_{k=1}^{N}\frac{det(C_{k})}{det(C)}$. Since the calculation of the determinant of large matrices can be computationally challenging, we use the equivalent estimate of the trace of the inverse of the Gram matrix, $\mathrm{Trace}(C^{-1})$, to calculate the average determinant ratio (see Kim et. al. for a full derivation) \cite{kim2018role}.

To understand the expected differences in stimulation-induced dynamics based on which region is actually being stimulated, we calculated two network control statistics: the \emph{persistent modal controllability} and the \emph{transient modal controllability}. Intuitively, the persistent (transient) controllability is high in nodes where the addition of energy will result in large perturbations to the slow (fast) modes of the system \cite{gu2015controllability}. Typically, modal controllability is computed from the eigenvector matrix $V = [v_{ij}]$ of the adjacency matrix $\mathbf{A}$. The $j^{\mathrm{th}}$ mode of the system is poorly controllable from node $i$ if the entry for $v_{ij}$ is small. Modal controllability is then calculated as $\phi_{i} = \sum_{j=1}^{N}(1 - \lambda^{2}_{j}(A))v^{2}_{ij}$. We adapt this discrete-time estimate to continuous-time by defining modal controllability to be $\phi_{i} = \sum_{j=1}^{N}(1 - (e^{\lambda_{j}(A)\delta  t})^{2})v^{2}_{ij}$. Here, $\delta t$ is the time step of the trajectory and $e^{\lambda_{j}(A)\delta  t}$ is the conversion from continuous to discrete eigenvalues of the system. Persistent (transient) modal controllability are computed in the same way, but using only the 10\% largest (smallest) eigenvalues of the system. We explore how this cutoff affects estimates of persistent and transient controllability in the Supplement. 

\subsection*{Random Forest Model}
Random forest models are constructed by averaging predictions over a large number of decision trees (here: 500), where each branch in the tree splits one of the predictors into two groups, the means of which are used as a predicted value for observations in each branch \cite{Liaw2002}. Splits are selected to reduce prediction error. Random forest models rely on bootstrapping data for each split, and a random selection of the variable to split on to avoid overfitting the data. Out-of-bag mean squared error is calculated as the prediction error of the samples that were not included in bootstrapped selection for each tree, and therefore are samples that the model has not been trained on \cite{Liaw2002}. For further details, see Supplemental Methods.

\subsection*{Data Sharing}
Code for simulations and select metrics is available at https://github.com/jastiso. Data will be made available upon request.

\section*{Acknowledgments}
This work was supported by the Alfred P. Sloan Foundation (DSB), the John D. and Catherine T. MacArthur Foundation (DSB), the NIH R01 NS099348 (DSB), and the ISI Foundation (DSB) in addition to the NSF BCS-1441502 (DSB) and NSF BCS-1631550 (DSB). Data collection was supported by the DARPA Restoring Active Memory (RAM) program (cooperative agreement N66001-14-2-4032). We thank Yousseff Ezzyat, Dan Rizzuto, Michael Kahana and other members of the Kahana lab for guidance and providing classifier output. As well as Michael Sperling and others at the Hospital at the University of Pennsylvania and Jefferson University Hospital for subject recruitment and stimulation monitoring. We thank Blackrock Microsystems for providing neural recording and stimulation equipment. The views, opinions, and/or findings contained in this material are those of the authors and should not be interpreted as representing the official views or policies of the Department of Defense or the U.S. Government. We are indebted to the patients and their families for their participation and support.

\section*{Author Contributions}
DSB, ANK, and JS designed analyses; JS analyzed data; FP, TM, and JS wrote code; AEK constructed DWI matrices; ANK preprocessed data; FP and TM developed control framework; JS and DSB wrote the manuscript. Other authors were involved in data collection and manuscript editing.

\section*{Author Declaration}
The authors declare no conflicts of interest.

\end{document}